\documentclass[10pt,twocolumn,letterpaper]{article}

\usepackage{cvpr}              

\usepackage[dvipsnames]{xcolor}

\newcommand{\ours}{iTACO\xspace}
\newcommand{\rsrd}{\textsc{RSRD}\xspace}
\newcommand{\artany}{\textsc{AA}\xspace}

\definecolor{cvprblue}{rgb}{0.21,0.49,0.74}
\usepackage[pagebackref,breaklinks,colorlinks,citecolor=cvprblue]{hyperref}

\usepackage[utf8]{inputenc} 
\usepackage[T1]{fontenc}    
\usepackage{url}            
\usepackage{booktabs}       
\usepackage{amsfonts}       
\usepackage{nicefrac}       
\usepackage{microtype}      
\usepackage{xcolor}         
\usepackage{graphicx}
\usepackage{multirow}
\usepackage{amsmath}
\usepackage{float}
\usepackage[capitalize]{cleveref}
\crefname{section}{Sec.}{Secs.}
\Crefname{section}{Section}{Sections}
\Crefname{table}{Table}{Tables}
\crefname{table}{Tab.}{Tabs.}
\usepackage{xcolor}

\def\paperID{258} 
\def\confName{3DV\xspace}
\def\confYear{2026\xspace}

\title{{\color{Orchid}\ours}: {\color{Orchid}I}nteractable Digital {\color{Orchid}T}wins of {\color{Orchid}A}rticulated {\color{Orchid}O}bjects\\from Casually Captured RGBD Videos}

\author{%
  \textbf{Weikun Peng$^{1}$, Jun Lv$^{2}$, Cewu Lu$^{2}$, Manolis Savva$^{1}$}\\
  $^1$Simon Fraser University
  $^2$Shanghai Jiao Tong University \\
  \texttt{\color{cyan} \href{https://3dlg-hcvc.github.io/video2articulation/}{3dlg-hcvc.github.io/video2articulation/}}
}

\begin{document}
\maketitle
\begin{abstract}
Articulated objects are prevalent in daily life. Interactable digital twins of such objects have numerous applications in embodied AI and robotics. Unfortunately, current methods to digitize articulated real-world objects require carefully captured data, preventing practical, scalable, and generalizable acquisition. We focus on motion analysis and part-level segmentation of an articulated object from a casually captured RGBD video shot with a hand-held camera. A casually captured video of an interaction with an articulated object is easy to obtain at scale using smartphones. However, this setting is challenging due to simultaneous object and camera motion and significant occlusions as the person interacts with the object. To tackle these challenges, we introduce \ours: a coarse-to-fine framework that infers joint parameters and segments movable parts of the object from a dynamic RGBD video. To evaluate our method under this new setting, we build a dataset of 784 videos containing 284 objects across 11 categories that is 20$\times$ larger than available in prior work. We then compare our approach with existing methods that also take video as input. Our experiments show that \ours outperforms existing articulated object digital twin methods on both synthetic and real casually captured RGBD videos.

\end{abstract}    
\begin{figure*}
\centering
\includegraphics[width=0.95\textwidth]{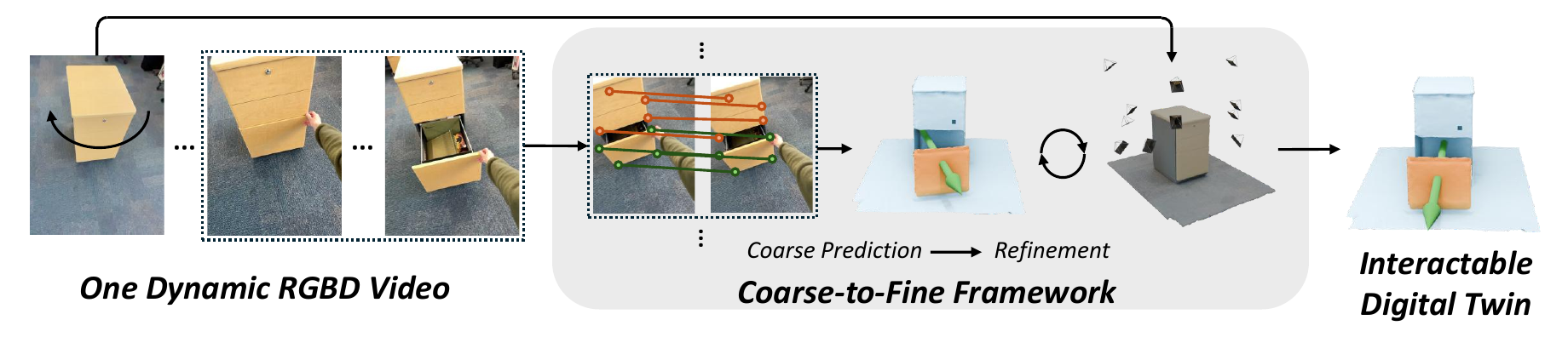}
\vspace{-4mm}
\caption{We propose \ours: a coarse-to-fine framework for building interactable digital twins of articulated objects from a casually captured RGBD video. Our pipeline first predicts coarse joint parameters and movable part segmentations. The initial estimates are then refined with gradient-based optimization.}
\label{fig:teaser}
\vspace{-4mm}
\end{figure*}

\section{Introduction}
\label{sec:intro}

Building digital twins of articulated objects has many applications in computer animation, embodied AI, and robotics. For example, scalable pipelines for building interactable digital twins benefit robotics research, by enabling convenient training and evaluation in simulation prior to real-world deployment. However, current approaches to building interactable replicas of articulated objects impose many constraints on the input data, preventing practical and efficient acquisition of articulated objects in realistic settings. Thus, researchers still largely rely on manual annotation of articulation joint parameters~\cite{torne2024rialto}.


Recent approaches to articulated object reconstruction and motion analysis can be categorized into three families. The first focuses on dynamic reconstruction of the shape of articulated objects~\cite{mu2021learning, yang2022banmo, tseng2022cla, wei2022self, song2024reacto}. Unfortunately, the final output is not interactable as it is typically a sequence of meshes instead of a kinematic structure with explicit links and joints. The second family infers joint parameters and builds part-level geometry simultaneously from two different states of the object~\cite{jiang2022ditto, jiayi2023paris, weng2024neural, liu2025building, lv2022sagci, nie2023structure}. However, this requires aligning observations between states, and even small errors in alignment can lead to dramatic drops in accuracy. The third family of approaches generates articulated objects leveraging large language and multimodal models~\cite{mandi2024real2code, le2024articulate}. At present, these approaches either rely on an external library for object retrieval or require data for fine-tuning. Retrieving objects is not easily scalable, while collecting data for fine-tuning is labor-intensive. Furthermore, generalization to real settings after fine-tuning on limited data is not easy.

By analyzing the issues of recent approaches, we summarize three desirable features for practical interactable digital twin acquisition: 1) \textbf{minimal input constraints}; 2) \textbf{explicit representation} of joint parameter and part-level geometry for easy integration into simulators; and 3) \textbf{generalization} across diverse object categories. To address these desiderata, we create digital twins of articulated objects from casually captured RGBD videos showing a human interacting with the object. For example, a human operator with a hand-held camera can first scan the object to reconstruct the object surface, and then record an interaction with the object. Such videos are easy to capture at scale and contain the necessary motion information to infer joint parameters. Compared to the problem setting in prior work, casually captured videos are far less constrained and easier to capture, the final output is easier to use for simulation, and the overall pipeline is scalable to real-world settings.


However, casually-captured video also brings more challenges. Since both the camera and the scene are dynamic, and the person occludes the object as they are manipulating it, decoupling camera motion and object part motion is challenging. Consequently, accurate estimation of joint parameters and movable part segmentation is difficult. To address these challenges, we develop \ours: a coarse-to-fine framework to build digital twins of articulated objects from dynamic RGBD videos, illustrated in \cref{fig:teaser}. Our pipeline first estimates joint parameters and a movable part segmentation via image feature matching. Then, a gradient-based optimization framework refines the initial estimates against a surface point cloud representation of the object acquired through 3D reconstruction. Our pipeline explicitly parameterizes articulation and geometric parameters, satisfying the second desideratum. Moreover, to aid generalization and satisfy the third desideratum, our pipeline neither relies on an external library nor requires additional data for fine-tuning. Instead, it only uses pretrained models.

Since our problem has not been previously addressed, we build a dataset for evaluation. We generate videos for 284 synthetic objects from 11 categories in PartNet-Mobility~\cite{Mo_2019_CVPR} and collect real RGBD videos. Our dataset contains 20$\times$ more objects compared to prior works~\cite{jiayi2023paris,liu2025building}, allowing a more robust evaluation of articulated object acquisition.
In summary, we make the following contributions: 
\begin{enumerate}
    \item We address building interactable digital twins of articulated objects from casually captured RGBD videos, a new problem setting that is more practical than prior work.
    \item We develop \ours: a coarse-to-fine framework for acquisition of digital twins of articulated objects.
    \item We collect a dataset with 20$\times$ more objects than prior datasets. Experiments show that \ours outperforms state-of-the-art methods on both synthetic and real articulated objects across different categories.
\end{enumerate}
\section{Related Work}
\label{sec:relate}
Our work is related to motion analysis and reconstruction of articulated objects, and dynamic scene understanding.

\subsection{Articulated Object Reconstruction}
Motion analysis and reconstruction of articulated objects is a long-standing task in computer vision and robotics. Previous motion analysis works focus on predicting joint parameters or part-level segmentation from an image or a single point cloud~\cite{wang2019shape2motion, Yan2019RPMNet, li2019category, weng2021captra, jiang2022opd, sun2023opdmulti, yu2024gamma}. These works require training on specific datasets, which limits generalization. Previous articulated object reconstruction works focus on reconstructing the shape of articulated objects~\cite{mu2021learning, yang2021lasr, yang2022banmo, wei2022self, song2024reacto}. The final output of these methods is the entire shape of the object, which is \textbf{not interactable}, in particular due to the absence of a kinematic structure with joints and articulation parameters that can be driven kinematically or through physics-based simulation. In contrast, we aim to build \textbf{interactable} digital twins of articulated objects, where inferring joint parameters and part-level geometry is essential.

Another line of work attempts to infer joint parameters and build part-level geometry of articulated objects simultaneously. One problem setting uses multi-view observations of different object states~\cite{jiang2022ditto, jiayi2023paris, weng2024neural, liu2025building}. These works implicitly assume all observations have been aligned to the same coordinates. In practice, this requires either manual alignment which is not scalable or algorithmic alignment which is error-prone. Other prior work uses robots to interact with the articulated object and reconstruct articulated objects from the interaction~\cite{lv2022sagci, nie2023structure, InterRecon23, wang2024articulated}. However, robots are expensive and may damage themselves or the environment during the interaction. Therefore, this is not a practically scalable path to build digital twins of articulated objects. Recently, researchers began exploring motion analysis from videos~\cite{kerr2024rsrd, werby2025articulated, zhou2025monomobility}. Robot-See-Robot-Do~\cite{kerr2024rsrd} recovers 3D motions of an articulated object from a monocular human demonstration video. Their problem setting is close to us, but requires the camera to be static. 

Other work uses large language and multimodal models to generate a code representation or kinematic graph from a single image of an articulated object~\cite{mandi2024real2code, le2024articulate, chen2024urdformer, jiayi2024singapo}. Leveraging vision-language models, Articulate-Anything can also take image or video as input~\cite{le2024articulate}. These works either rely on existing mesh libraries to retrieve object part meshes or require extra data for fine-tuning foundation models. Thus, at the current stage, this line of work still suffers from out-of-distribution issues, limiting generalization and scalability. In contrast, our method does not rely on object retrieval or fine-tuning, and is inherently more general and scalable.

In computer animation, automatic skeletal rigging has been studied for decades~\cite{Tagliasacchi2009curve, Baran2007auto, yan2016erosion, Tzionas2016reconstructing, yan2018voxel, xu2020rignet, li2021nbs, xu2022morig, sun2025drive, Song2025magicarticulate, deng2025anymate, liu2025riganything, zhang2025unirig}. However, the skeleton model is designed to animate humans and animals, whose links are non-rigid and joints are mainly ball joints. For rigid articulated objects such as drawers and microwaves, most joints are prismatic joints and revolute joints. Thus, it is suitable to represent rigid articulated objects using skeleton models.

\subsection{Dynamic Scene Reconstruction}
Dynamic scene reconstruction is a challenging task in computer vision. One line of research describes a dynamic scene with a deformation field. Signed distance functions, and later neural radiance fields and 3D gaussian splatting are commonly used~\cite{newcombe2015dynamicfusion, bozic2020deepdeform, slavcheva2018sobolevfusion, pumarola2020d, duan:2024:4drotorgs, yang2023gs4d, luiten2023dynamic, Wu_2024_CVPR, som2024, xu2024das3r, wang2025gflow}. These works achieve impressive results in shape reconstruction and novel view rendering of dynamic scenes, but it is difficult to transform the deformation field to an explicit motion representation, such as a rigid transformation matrix, which is necessary for our problem.

Another line of approaches attempts to reconstruct dynamic scenes from monocular depth estimation and optical flow. Robust-CVD proposes to first estimate the depth of each frame and then optimize camera poses and depth jointly~\cite{kopf2021rcvd}. CasualSAM leverages optical flow to train a depth and movement prediction network and optimize camera parameters~\cite{CasualSaM}. MegaSaM proposes an optimization framework for dynamic scene reconstruction based on DROID-SLAM~\cite{li2024_MegaSaM, teed2021droid}. Recently, DUSt3R~\cite{dust3r_cvpr24} provides a powerful foundation model for static scene reconstruction. There are several recent extensions to DUSt3R. MonST3R fine-tunes DUSt3R on dynamic scene datasets to recover dynamic 3D scenes~\cite{zhang2024monst3r}. CUT3R introduces a continuous 3D perception model for online dense 3D reconstruction~\cite{cut3r}. Easi3R identifies moving objects in the video from DUSt3R's attention maps~\cite{chen2025easi3r}. Aether proposes a unified model for reconstruction, planning, and prediction in dynamic scenes~\cite{aether}. Our coarse prediction pipeline directly benefits from MonST3R~\cite{zhang2024monst3r}, and our refinement framework is inspired by the optimization framework in CasualSAM~\cite{CasualSaM} and MegaSaM~\cite{li2024_MegaSaM}.

A key sub-task in dynamic scene understanding is segmenting the `moving map' (mask image indicating which pixels are moving in the scene at each frame in the video). Recent approaches are mainly based on optical flow or point tracking~\cite{yang2021selfsupervised, meunier2023em, meunier2023unsupervised, xie2024flowsam, karazija24learning, golisabour2024romo, huang2025segmentmotionvideos}. However, we find that point tracking methods struggle with textureless objects and are sensitive to point locations (see supplement). Therefore, we do not use these approaches.
\section{Method}
\label{sec:method}

\subsection{Problem Formulation}
We define the interaction video as $\mathcal{V} = \{\mathcal{I}_t\}_{t=0}^T$, where $T$ denotes the video length and each frame $\mathcal{I}_t \in \mathbb{R}^{H\times W \times 4}$ is an RGBD image with height $H$ and width $W$.
An object surface point cloud $\mathcal{P^O}$ is assumed as input from a surface reconstruction stage prior to the start of interaction, as there are numerous methods for static surface reconstruction. The state of the object when reconstructing the surface is the initial state when recording the interaction video, and in practice can be captured in the same video right before the interaction begins.

We assume the video is casually captured with hand-held devices such as smartphones. Therefore, we cannot assume that the camera remains stationary during recording, nor do we assume that the camera trajectory can be easily obtained. We only assume the first video frame is aligned to the object surface coordinate system. In practice, this alignment can be achieved via point cloud registration or structure from motion. Normally, a human operator manipulates one joint at a time. Thus, we also assume only one part of the object is moving in the video. For multi-joint articulated objects, we can build each joint and part sequentially with consecutive interaction videos.

The output includes two key components: 1) joint parameters $\mathcal{J} \in \mathbb{R}^{7}$, including a binary value to indicate the type of joint (e.g., revolute or prismatic), as well as two triplets to indicate the position and orientation of the joint axis; and 2) movable parts of the object. Since it is challenging to segment movable parts on the 3D objects directly, we instead estimate a moving map for each video frame $\mathcal{M}_t = \{0,1\} \in \mathbb{R}^{H \times W}$, which indicates the moving parts of the object in the video. Then, we unproject the moving map to 3D space with depth values to identify movable parts of the object. In addition, we also need to estimate both camera pose $p^\text{cam}_t \in SE(3)$ and the joint state $s_t \in \mathbb{R}$ at each frame.



\begin{figure*}
\centering
\includegraphics[width=0.98\textwidth]{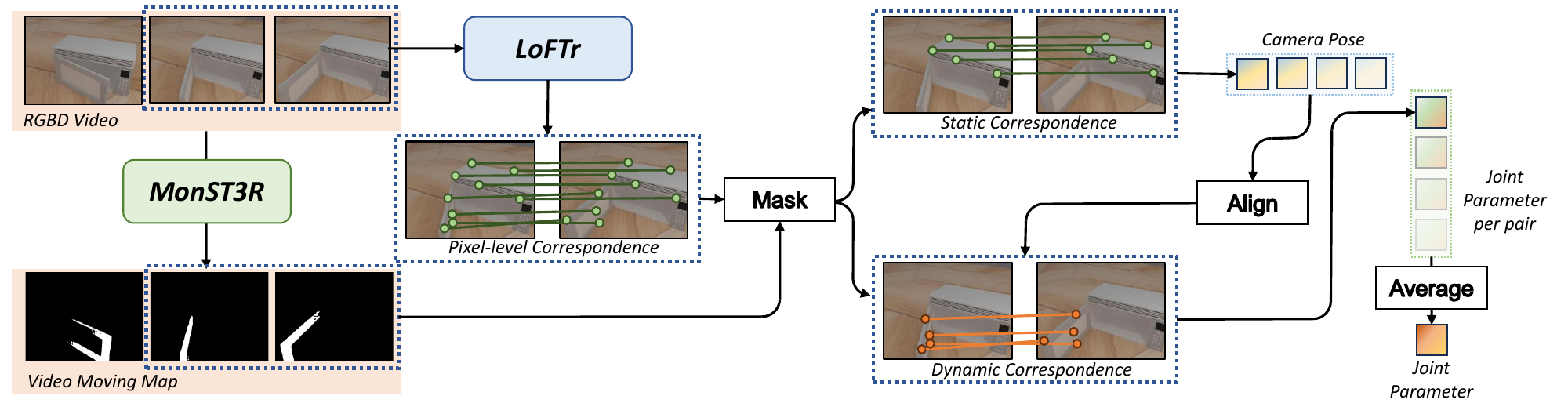}
\caption{An overview of our coarse prediction pipeline. We first use feature matching in the static regions to estimate relative camera poses and align all observations to the same coordinate. Then, we compute the rigid transformation using feature matching in the dynamic regions to estimate joint parameters. Finally, we average out all the results to produce a joint parameter estimation.}
\label{fig:coarse prediction}
\vspace{-4mm}
\end{figure*}

\subsection{Coarse Prediction}

We first design a coarse prediction pipeline as shown in \cref{fig:coarse prediction}. The idea is to decouple the problem by firstly finding the moving part $\mathcal{M}_t$ of each frame of the video $\mathcal{V}$. Once we identify the dynamic region and static region in the video frames, we can use point motion in the static region to estimate the camera pose ${p^\text{cam}_t}$ at each frame. We align all the observations to the same coordinate, and compute rigid transformations of movable parts of the object using points motion in the dynamic region to estimate joint parameters $\mathcal{J}$ and joint state $\{s_t\}_{t=0}^T$. 


In practice, we use MonST3R~\cite{zhang2024monst3r} to provide a rough moving map $\mathcal{M}_t$ of each video frame. Then we unproject pixels in $\mathcal{I}_t$ to 3D point cloud $\mathcal{P}_t$ with camera intrinsics, and use LoFTr~\cite{sun2021loftr} to compute pixel-level feature matching between two video frames. We use feature matching in static regions identified by $\mathcal{M}_t$ to compute relative camera poses. We align all observations to the same camera coordinate system with the relative camera poses of each frame. Then, we use feature matching in the dynamic regions to compute a rigid transformation of the movable parts of the object. We can further estimate joint parameters from this rigid transformation. We estimate joint parameters from multiple video frame pairs and average the results as the coarse prediction of the joint parameters and states. 

Through the coarse prediction stage, the proposed pipeline obtains rough results for moving map segmentation and joint parameters. However, this method has two disadvantages: 1) Since the moving map is inaccurate, camera poses and joint parameters estimation are inaccurate. 2) Feature matching provides a reliable prediction only when two video frames are visually similar. In other words, the joint states do not change much, and the point motion in dynamic regions will be minimal. In this situation, it's hard to tell which joint types lead to this kind of point motion. Joint type prediction will be inaccurate as well.

\subsection{Refinement}

\begin{figure*}
\centering
\includegraphics[width=0.95\textwidth]{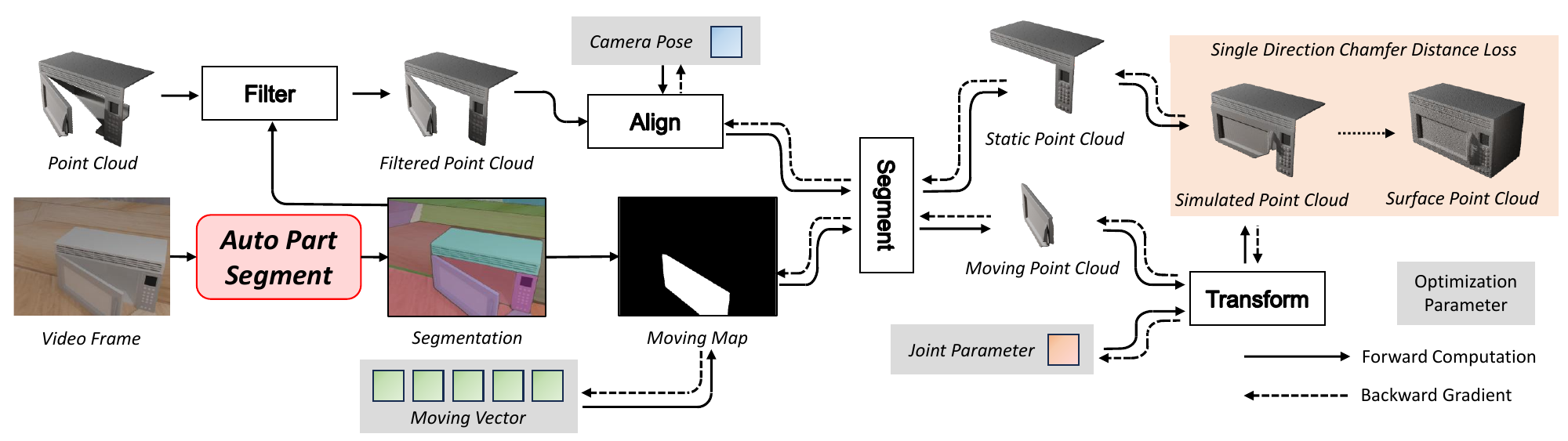}
\caption{An overview of our refinement pipeline. We transform observations in the video back to the initial stage with camera poses and joint parameters. We then compute the chamfer distance from the transformed observation to the object surface as a loss function and optimize relevant parameters.}
\label{fig:refinement}
\vspace{-4mm}
\end{figure*}

We refine the initial estimation results through gradient-based optimization. This phase is based on a simple idea: if the estimates of the joint parameters and camera poses are accurate, the object states observed in the video should transform back to the initial state. In other words, the transformed point cloud should align with the given object surface point cloud at the initial state. An overview of our pipeline is shown in \cref{fig:refinement}.

Concretely, we divide the optimization target into static part and moving part. For the static part, we transform the point cloud $\mathcal{P}_t$ to the same coordinate with camera poses $p_t^\text{cam}$. We compute a \textbf{single-directional Chamfer distance} to the given object surface point cloud $\mathcal{P^O}$ in \cref{eq:static chamfer}.

\begin{equation}
    \mathcal{L}_{\text{static}} = \frac{1}{T}\sum_{t=1}^T (1 - \mathcal{M}_t)\cdot \text{Chamfer}(p_t^\text{cam} \mathcal{P}_t, \mathcal{P^O})
    \label{eq:static chamfer}
\end{equation}

To compute the Chamfer distance of the moving part, we first compute the transformation matrix $k_t$ for the moving part with $\mathcal{J}$ and $s_t$. Then, we transform $\mathcal{P}_t$ with camera pose $p_t^\text{cam}$ and $k_t$. Then, we compute the \textbf{single-directional Chamfer distance} to $\mathcal{P^O}$ in \cref{eq:dynamic chamfer}.

\begin{equation}
    \mathcal{L}_{\text{dynamic}} = \frac{1}{T}\sum_{t=1}^T \mathcal{M}_t\cdot \text{Chamfer}(k_tp_t^\text{cam} \mathcal{P}_t, \mathcal{P^O})
    \label{eq:dynamic chamfer}
\end{equation}

Combining these two equations produces the final optimization objective $\mathcal{L} = \mathcal{L}_{\text{static}} + \mathcal{L}_{\text{dynamic}}$. Note that this computation graph is differentiable. Thus, we can use gradient-based optimization methods to optimize joint parameters, camera poses, and the moving map.

However, this na\"ive implementation has two problems. Firstly, it does not deal with newly observed parts of the object in the video. A newly observed part is a region that is initially hidden but becomes visible later (e.g., interior of a microwave). This region does not correspond with the object surface point cloud. Consequently, the Chamfer Distance is unreliable for point cloud alignment. Secondly, this implementation is quite inefficient. This is due to the need to optimize the probability of each pixel belonging to the moving part of each frame, which contains $T\times H\times W$ parameters in total.
Considering each pixel individually is not efficient as there are millions of pixels in each frame.

To tackle the second problem our automatic part segmentation module aggregates pixels into object parts. This is more efficient and effective as the number of parts is much smaller than the number of pixels, reducing optimization complexity. Similar to Segment-Anything~\cite{kirillov2023segment}, automatic part segmentation segments the first video frame into several part regions, and then tracks these regions throughout the video using SAM2~\cite{ravi2024sam2}. Subsequently, instead of optimizing $\mathcal{M}_t$ on each pixel, we optimize the probability of being a moving part for each tracked segment. We define a vector $v\in (0,1)^{D\times 1}$ to represent the probability of each segment being a moving part, where $D$ is the number of segments. To construct $\mathcal{M}_t$ for a frame, we multiply each probability value by the corresponding mask of the part. Finally, we sum the results across different parts to construct the moving map. We leverage an existing \href{https://github.com/zrporz/AutoSeg-SAM2}{implementation} of object part segmentation for videos.

The above automatic part segmentation module is also used to identify newly observed parts, thus helping to alleviate the first problem with the na\"ive implementation. If a part does not appear in the first frame, then it is classified as a newly observed part. By filtering out newly observed parts in the video, we reduce the spurious correspondence issue when computing the Chamfer Distance.

Since our pipeline only estimates the moving map of the input video, we need to find the movable part of the object on $\mathcal{P^O}$. We unproject the pixels in the moving region in the moving map of the first frame $\mathcal{M}_0$. Then, we select points on $\mathcal{P^O}$ which are close to the unprotected moving pixels to be the movable part of the object. We use NKSR~\cite{huang2023nksr} to build meshes of both the static and the movable parts of the object.

\begin{table*}
    \centering
    \renewcommand{\arraystretch}{1.2}
    \begin{tabular}{c l c c c c c}
        \toprule
        Joint & Methods
        & Axis(rad)$\downarrow$ & Position(m)$\downarrow$ & Type(\%)$\downarrow$ & State(rad or m)$\downarrow$ & Failure(\%)$\downarrow$ \\
        \midrule
          \multirow{6}{*}{Revolute}
          & \artany~\cite{le2024articulate}           & 0.82$\pm$0.79 & 0.81$\pm$0.40 & 40.90 & N.A. & 38.63 \\
          & \rsrd~\cite{kerr2024rsrd}                          & 1.16$\pm$0.52 & 1.18$\pm$1.21 & 88.88 & 1.05$\pm$0.59 & 46.66 \\
              & \ours (Ours)                        & \textbf{0.32$\pm$0.56} & \textbf{0.13$\pm$0.25} & \textbf{15.90} & \textbf{0.25$\pm$0.46} & 6.81 \\
          & \ours w/o Refine           & 1.03$\pm$0.67  & 0.24$\pm$0.25 & 68.18 & 0.59$\pm$0.47 & 4.54 \\
          & \ours w/o Coarse    & 1.26$\pm$0.29  & 0.34$\pm$0.26 & 77.27 & 0.69$\pm$0.45 & \textbf{2.27} \\
          & \ours w/o Segment    & 0.36$\pm$0.55  & 0.17$\pm$0.24 & 18.18  & 0.33$\pm$0.44 & 6.81 \\
        \cmidrule{1-7}
          \multirow{6}{*}{Prismatic}
              & \artany~\cite{le2024articulate}           & 0.92$\pm$0.78 & N.A. & 50.00 & N.A. & 46.66 \\
          & \rsrd~\cite{kerr2024rsrd}                        & 1.27$\pm$0.44 & N.A. & 40.00 & 0.63$\pm$0.41 & 36.66 \\
              & \ours                         & \textbf{0.24$\pm$0.33} & N.A. & 10.34 & 0.08$\pm$0.22 & \textbf{0} \\
          & \ours w/o Refine           & 0.28$\pm$0.40  & N.A. & \textbf{6.89} & 0.16$\pm$0.16 & \textbf{0} \\
          & \ours w/o Coarse    & 0.42$\pm$0.11  & N.A. & \textbf{6.89} & 0.08$\pm$0.09 & \textbf{0} \\
          & \ours w/o Segment    & 0.30$\pm$0.41  & N.A. & 13.79 & \textbf{0.07$\pm$0.07} & \textbf{0} \\
        \bottomrule
    \end{tabular}
    \caption{Evaluation of articulation parameter estimation on S-Dataset. We report the mean and standard deviation across all the test cases within the dataset. \ours performs best across most metrics, and ablations show the value of each component.}
    \label{tab:main_results}
    \vspace{-15pt}
\end{table*}

\section{Experiments}
\label{sec:exp}

We aim to answer the following questions with our experiments:
1) How well can we estimate joint parameters?
2) How well can we acquire the geometry of the articulated object?
3) How do different modules and strategies of our pipeline contribute to the final results? and
4) Can our method work in real-world scenarios?

\subsection{Experiment Setup}

\paragraph{Dataset.}
At present, there is no benchmark for evaluating joint parameter estimation from dynamic RGBD videos. Therefore, we construct a synthetic dataset for evaluation. We select objects from 11 categories in the PartNet-Mobility dataset~\cite{Mo_2019_CVPR}. We simulate interaction between human hands and these objects using the SAPIEN simulator~\cite{Xiang_2020_SAPIEN}. For each interaction, we generate two videos shot from two different viewpoints. Camera trajectories are generated by moving the camera towards a new target pose every 7 - 10 timesteps in the simulator. The target pose is sampled from a Gaussian distribution, where the mean value is the current camera pose and the standard deviation is set to ensure the camera can consistently look at the object. To get the object surface point cloud, we unproject the depth maps from 24 viewpoints to 3D and fuse the point cloud. For Robot-See-Robot-Do, we generate 100 views because it requires producing a NeRF and 3D Gaussian splat~\cite{kerr2024rsrd}. Please refer to the supplement for more details about the dataset construction. Since Robot-See-Robot-Do requires manually cropping and clustering 3D gaussian splats which is labor-intensive, we split the dataset into two parts. We randomly sampled 10\% of test videos from our dataset to construct a smaller dataset named S-Dataset, containing 73 test videos, while L-Dataset contains the remaining videos in the original dataset.

\paragraph{Baselines.}
We select two recent works on modeling interactable digital twins of articulated objects from video input:
\begin{itemize}
    \item \textbf{Robot-See-Robot-Do} (\rsrd)~\cite{kerr2024rsrd}: a differentiable rendering pipeline that recovers 3D part motion from human demonstration videos. We follow the pipeline to first reconstruct and segment the 3D Gaussian splats of the object. Then, the input video is used to estimate the SE3 transformation of the 3D Gaussians for each part at each frame via differentiable rendering. For a fair comparison, we modify their tracking pipeline to exploit ground truth depth maps of the input video. We select the group of 3D Gaussians with the longest moving distance to be the moving part of the object. Since this pipeline does not include a geometry reconstruction module, we export 3D Gaussians and reconstruct a mesh using NKSR~\cite{huang2023nksr} for geometry reconstruction. 
    \item \textbf{Articulate Anything} (\artany)~\cite{le2024articulate}: a system that retrieves object meshes from a library and estimates link and joint parameters from text, image, or video input. We follow the original work and use the Partnet-Mobility dataset as the object mesh library, which contains all the ground truth object meshes for our test cases, and use Gemini Flash-1.5~\cite{team2023gemini} for the VLM module.
\end{itemize}

\paragraph{Metrics.}
We use a suite of metrics to evaluate both geometric reconstruction and articulation parameter estimation.
Geometric accuracy is evaluated using Chamfer Distance between ground truth and estimated geometry for the whole object (CD-w), for the movable part (CD-m) and for the static part (CD-s).
Articulation parameter estimation is evaluated using joint type classification error (percentage), joint axis error (radians), joint position error for revolute joints (meters), and joint state error (radians for revolute joints and meters for prismatic joints). If the method crashes or retrieves incorrect objects (for retrieval methods), we classify the output as a failure and assign $\frac{\pi}{2}$ error for joint axis and revolute joint state, and error equal to 1 for other metrics.

\subsection{Articulation Parameter Estimation Results}

\begin{table}
    \centering
    \renewcommand{\arraystretch}{1.2}
    \resizebox{\columnwidth}{!}{
    \begin{tabular}{l c c c} 
        \toprule
        Methods & CD-w$\downarrow$ & CD-m$\downarrow$ & CD-s$\downarrow$ \\
        \midrule
         \artany~\cite{le2024articulate} & 0.11$\pm$0.22 & 0.59$\pm$0.73 & 0.07$\pm$0.18 \\
         \rsrd~\cite{kerr2024rsrd} & 3.39$\pm$21.50 & 0.60$\pm$0.60 & 0.17$\pm$0.44 \\
         \ours (Ours) & \textbf{0.01$\pm$0.01} & \textbf{0.13$\pm$0.26} & \textbf{0.06$\pm$0.19} \\
         \ours w/o Refine & \textbf{0.01$\pm$0.01} & 0.40$\pm$0.44 & 0.35$\pm$0.46 \\
         \ours w/o Coarse & \textbf{0.01$\pm$0.01} & 0.19$\pm$0.22 & \textbf{0.06$\pm$0.16} \\
         \ours w/o Segment & \textbf{0.01$\pm$0.01} & 0.39$\pm$0.44 & 0.36$\pm$0.46 \\
        \bottomrule
    \end{tabular}
    }
    \caption{Geometry evaluation on S-Dataset. We report the mean and standard deviation across test cases within the dataset. \ours achieves the lowest reconstruction error among all aspects.}
    \label{tab:geometry_eval}
    \vspace{-15pt}
\end{table}

The quantitative results on S-Dataset are in \cref{tab:main_results}, and qualitative results are in \cref{fig:sim_qualitative_result}. Results on L-Dataset are in the supplement.
Our \ours method outperforms or is comparable to state-of-the-art methods along most metrics. \artany struggles with retrieving the correct object meshes. Most failure cases are due to retrieval errors, which is surprising since the retrieval library contains all the ground truth object meshes. For example, in \cref{fig:sim_qualitative_result} \artany retrieves a lamp in the Stapler test case. In addition, the link placement and affordance detection modules appear to not be robust either, showing the challenge of our problem setting. \artany selects the wheel of the table to be a movable part, and places the lid of the USB stick below the main body.

At the same time, \rsrd cannot represent textureless objects accurately, likely due to limitations in using 3D Gaussian splatting. Therefore, it often fails to correctly crop and cluster 3D Gaussians into parts, even though human operators participate in this step. Most importantly, \rsrd cannot handle camera motion. Thus, its movable part segmentation deviates from the ground truth significantly. In contrast, \ours produces more accurate object articulations.

\begin{figure*}
\centering
\includegraphics[width=0.9\textwidth]{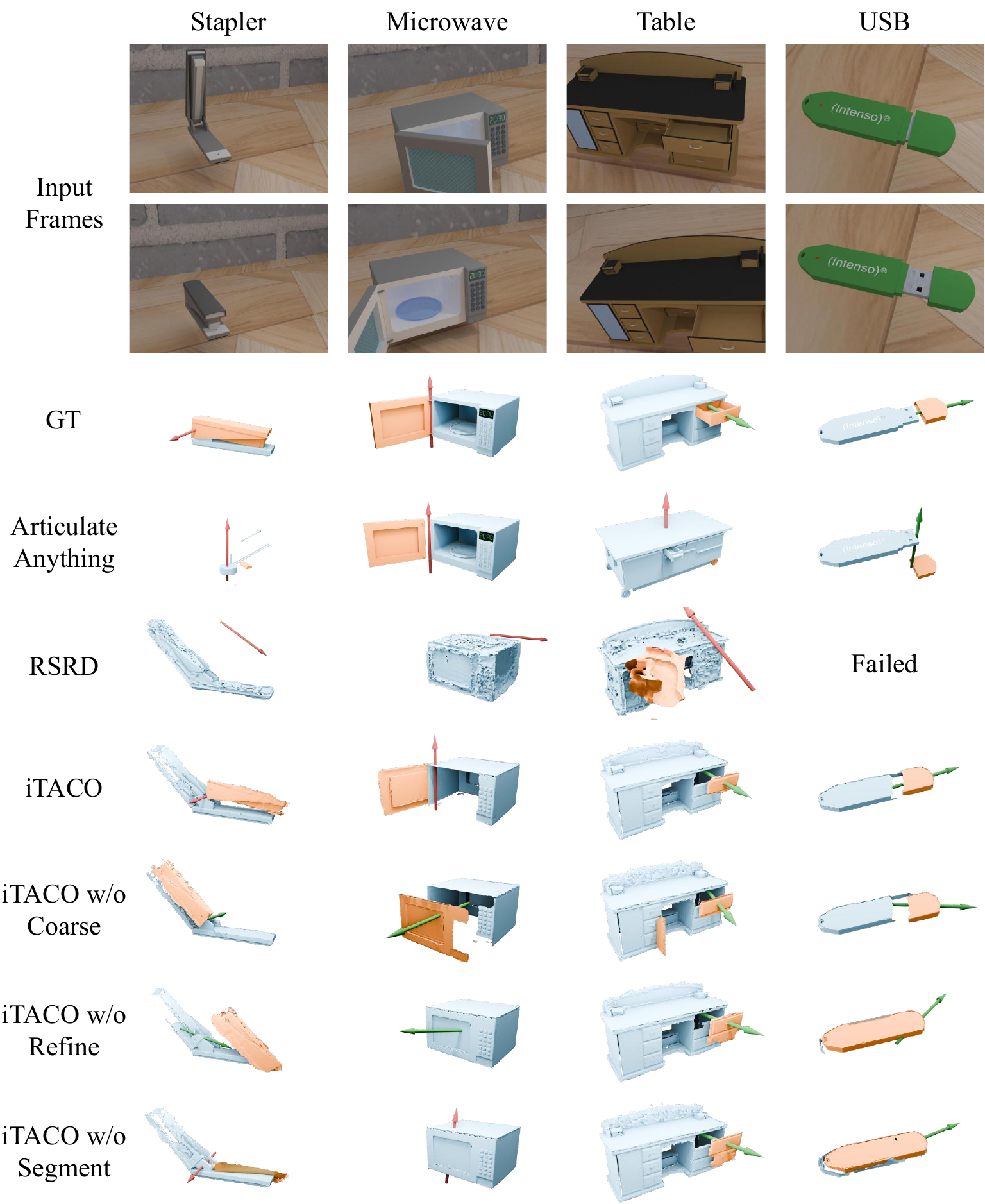}
\caption{Qualitative results on synthetic data. The moving parts of the object are shown in orange and static parts are in blue. Prismatic joints are green and revolute joints are red. To illustrate the joint state prediction results, we render the articulated object at the end state.}
\label{fig:sim_qualitative_result}
\vspace{-15pt}
\end{figure*}

\begin{figure*}
\centering
    \includegraphics[width=0.98\textwidth]{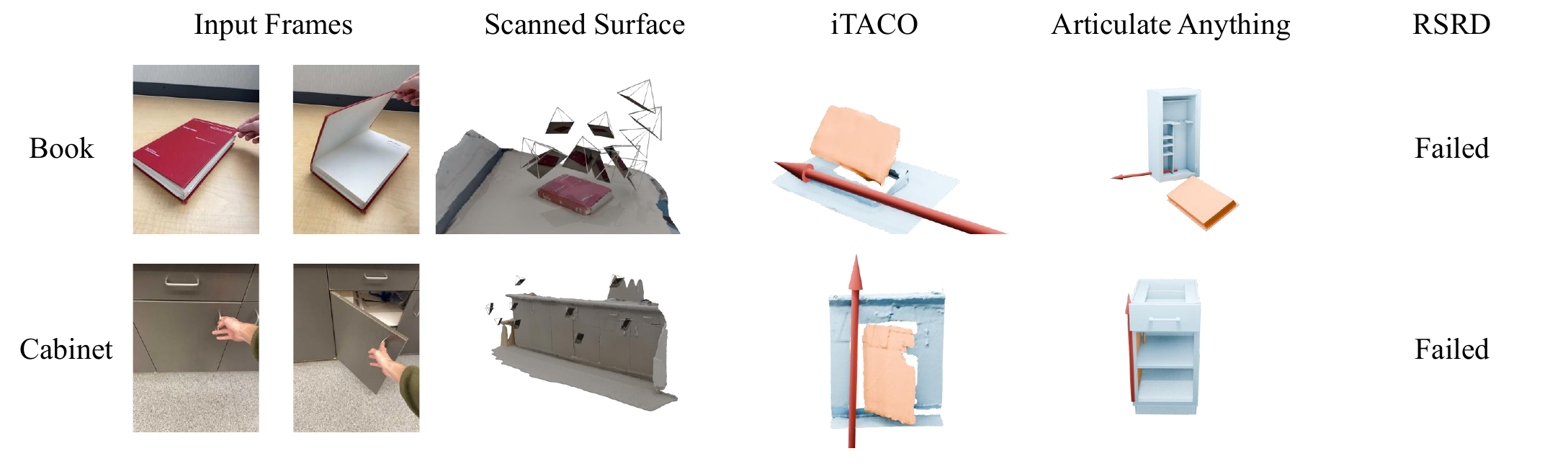}
    \caption{Qualitative results on real data. We find that Articulate Anything struggles with objects that do not exist in the mesh library, such as books. \rsrd does not work well on textureless objects such as the cabinet example.}
    \label{fig:real_qualitative_result}
\vspace{-4mm}
\end{figure*}

\subsection{Geometry Estimation Results}

We evaluate the geometry of our outputs using the Chamfer Distance to the ground truth. We sample 10000 points from the reconstructed mesh and the ground truth mesh, and compute the Chamfer Distance between them. \cref{tab:geometry_eval} shows the results on S-Dataset. Results on L-Dataset are in the supplement. Due to retrieval errors, link placement errors, and affordance detection errors, \artany has a high reconstruction error, particularly on movable parts. \rsrd performs even worse, struggling with clustering 3D Gaussian splats into different parts.

\subsection{Ablations}

We design three ablations of our method. To demonstrate the importance of the coarse prediction module, we remove it from our pipeline and randomly initialize the parameters of the refinement module, named \textbf{\ours w/o Coarse}. To illustrate the importance of the refinement module, we remove the refinement module and evaluate coarse prediction results directly, named \textbf{\ours w/o Refine}. Lastly, to demonstrate the advantage of the automatic part segmentation strategy for the refinement process, we na\"ively optimize the probability of being a moving part for each pixel at each video frame, named \textbf{\ours w/o Segment}. 

In \cref{tab:main_results} we find that without refinement, coarse prediction will fail on joint type prediction for more than half of the test cases, particularly on revolute joints. As mentioned in \cref{sec:method}, when two video frames are similar it is hard to distinguish revolute joints from prismatic joints. In \cref{fig:sim_qualitative_result}, we also see that the coarse prediction module misclassifies the Stapler and Microwave joints. Without the refinement module it is much harder to correctly segment the movable and static parts, as shown in \cref{tab:geometry_eval}. Also, refinement cannot work effectively from a random initialization in the ablation without coarse prediction, as shown quantitatively in \cref{tab:geometry_eval,tab:main_results} and qualitatively in \cref{fig:sim_qualitative_result}.

Without automatic part segmentation, the joint prediction performance degrades as the refinement module suffers from incorrect point correspondences. For instance, the Microwave test case in \cref{fig:sim_qualitative_result} fails. The segmentation also deviates from the correct segmentation as shown in \cref{tab:geometry_eval}. In \cref{fig:sim_qualitative_result}, we also see that without automatic part segmentation, our method has difficulty segmenting the moving parts and static parts accurately. This illustrates the importance of automatic part segmentation in optimizing joint parameters and the moving map from the video.

\subsection{Real-World Qualitative Evaluation}

We also evaluate our methods and baselines on real data. We use an iPhone 12 Pro with LiDAR camera to capture data. We use Polycam~\cite{polycam} to reconstruct the object surface and Record3D~\cite{record3d} to record the RGBD video of the human interaction with the object. To improve depth map quality, we use Prompt Depth Anything~\cite{lin2025prompting} to scale up the original depth map. We compute the image feature matching between the first frame of the video and all the images of the reconstruction stage with LoFTr~\cite{sun2021loftr}. We select the image pair with the most reliable feature matching, and compute the rigid transformation between corresponding points of the image pair to align the first video frame to the object surface coordinate system. Real videos include human hands, which interfere with MonST3R predictions for the moving map. Thus, we use Grounded SAM 2~\cite{GroundedSAM2, ren2024grounded} to mask hands with the text prompt: ``hands and arms''.

From the qualitative results shown in \cref{fig:real_qualitative_result}, we see that \artany fails on the Book test case due to the absence of similar meshes in the mesh library. In the Cabinet test case, \artany retrieves reasonable meshes but predicts the wrong joint rotation direction, as the door rotates into the cabinet. \rsrd fails to produce valid results due to the clustering and tracking issues on 3D gaussian. This demonstrates our method can work on real data and outperform current methods in the challenging setting of casually captured real-world RGBD video inputs.

\section{Conclusion}
\label{sec:conclusion}

We introduced a new problem setting for reconstructing articulated objects from casually captured RGBD videos. Our problem setting is more practical than prior works, bringing new challenges to current methods. We then developed \ours: a coarse-to-fine framework for this problem. To evaluate our method, we collected a new dataset that contains 20$\times$ more objects than previous benchmarks. Our experiments show our method outperforms baselines, providing a practical and robust approach to acquiring digital twins of articulated objects. Nonetheless, there is still space for improvement. Better dynamic scene understanding approaches and video segmentation models can help improve our method. In addition, our pipeline does not reconstruct the interior of the object. Reconstructing the interior of the object is an interesting direction for future work.

\vspace{1em}
\noindent\textbf{Acknowledgements.}
This work was funded in part by a Canada Research Chair, NSERC Discovery Grant, and enabled by support from the Digital Research Alliance of Canada. The authors would like to thank Jiayi Liu, Xingguang Yan, Austin T. Wang, Hou In Ivan Tam, Morteza Badali for valuable discussions, and Yi Shi for proofreading.

{
    \small
    \bibliographystyle{ieeenat_fullname}
    \bibliography{main}
}

\newpage
\appendix
\renewcommand\thesection{\Alph{section}}
\renewcommand\thefigure{\thesection.\arabic{figure}}
\renewcommand\thetable{\thesection.\arabic{table}}
\setcounter{figure}{0}
\setcounter{table}{0}
\setcounter{equation}{0}
\clearpage
\setcounter{page}{1}
\maketitlesupplementary

\section{Implementation Details}
\label{sec:implementation}
In this section we introduce more details about implementation of our method and baselines.

\subsection{Method}
\label{subsec:our implementation}

\paragraph{Coarse Prediction.}
Since the input video usually contains more than 100 frames, we sample the input video to around 20 frames. We first send these sampled frames to MonST3R~\cite{zhang2024monst3r} to compute the moving map of each frame. Then, we select pairs of frames to compute the feature matching using LoFTr~\cite{sun2021loftr}. Not all feature matches are accurate, so we only consider feature matches that have a confidence value larger than 0.95 to be valid. We first use feature matching within the static regions to compute the relative camera poses $p_{t,t-1}^\text{cam}$ from frame $t$ to $t-1$ sequentially. We use these relative camera poses to align all the frames to the first camera coordinate. 

Then, we proceed to estimate joint parameters. We select frame pairs within 3 time steps to ensure the accuracy of feature matching. Due to the inaccuracy of moving map and feature matching, we filter out frame pairs that contain fewer than 80 valid feature matches. For the rest of the frame pairs, we estimate joint parameters for the revolute joint and the prismatic joint separately. For each joint type, we run RANSAC~\cite{RANSAC} for 50 iterations to select the joint parameters estimation that best fits the observation. Based on the estimated joint parameters, we can transform the feature points $x_t$ to expect position $x'_{t+d}$ at time step $t+d$. By computing the distance between the expected position $x'_{t+d}$ and the observed position $x_{t+d}$, we can evaluate how well this set of joint parameters describes the observation. If the distance for the revolute joint is lower, this frame pair votes for revolute, and vice versa. After estimating joint parameters for all the frame pairs, we compute the mean value for the joint axis and position for both joint types. We choose the joint type that gets the most votes as the predicted joint type for coarse prediction.

\paragraph{Refinement.}
Though we predict the joint type in coarse prediction, we still retain the prediction results of both joint types from coarse prediction and initialize the refinement module with those values separately. For moving vector initialization, we compute the intersection area between each part and the moving map from MonST3R~\cite{zhang2024monst3r}, and then normalize it by the area of each part. We run the forward computation described in the main paper. But computing the Chamfer Distance will be very slow if the number of points is very high. Therefore, for each frame, we randomly sample 50\% of points from the whole observation to compute the Chamfer Distance. The gradient will not be affected since we are optimizing the moving vector instead of the probability of being a moving part of each pixel at each frame. This strategy significantly improves the optimization speed without losing prediction accuracy. We concurrently optimize the joint prediction for both joint types for 400 iterations, using Adam optimizer~\cite{kingma2014adam} and 5e-3 learning rate. 

After optimization, we have two sets of optimized joint parameters, and we need to choose the joint type. We first classify the test case as a prismatic joint if the distance between the object surface point cloud and the joint axis is larger than 0.1 meters. For the remaining cases, we compare the lowest chamfer distance of these two joint types and select the joint type that has the lower Chamfer Distance. These are the final joint type prediction results for our method. Finally, we find pixels in the moving map whose value is larger than 0.7 to be the moving points. We project these points back to 3D and find points in the object surface point cloud whose distance to the moving points is less than 0.03 meters to be the movable parts of the object. The other points in the object surface point cloud are classified as static parts of the object. We reconstruct meshes using NKSR~\cite{huang2023nksr}.

\paragraph{Hardware Requirements.}
Our method requires a GPU with at least 24GB of CUDA memory. For one test video, our coarse prediction module needs around 2 minutes to run MonST3R to get moving map of the video and another 30 seconds to estimate joint parameters, using RTX4090 GPU. For the refinement, we first need to run automatic part segmentation on the test video, which takes 4-8 minutes. Then, we need to run optimization for the two joint types separately. Thus, we recommend using multiple GPUs for this stage. For one joint type, our optimization needs 3-5 minutes to run on an RTX4090 GPU, or 8-15 minutes on an RTX A5000 GPU. Finally, for mesh reconstruction, to get better reconstruction quality, we use an RTX A6000 GPU to reconstruct the mesh, since it has 48GB of CUDA memory to allow more points for reconstruction.

\subsection{Baselines}
\label{subsec:baseline implementation}

\paragraph{Robot-See-Robot-Do (\rsrd).}
We follow the original pipeline of \rsrd to run the experiments on our dataset. \rsrd will estimate the motion of each part at each frame. Based on the part motion, we select the motion of the part that has the largest translation distance throughout the video to compute the joint parameters. For geometry reconstruction, we export the gaussian splats from the dig model in \rsrd pipeline~\cite{kerr2024rsrd} and build mesh using NKSR~\cite{huang2023nksr}.

\paragraph{Articulate Anything (\artany).}
We follow the original pipeline of \artany to run the experiments on our dataset. \artany will retrieve top-k objects from the Partnet-Mobility dataset based on the input video. Then, it selects the best one from top-k candidates~\cite{le2024articulate}. When evaluating its results, we consider them to be a failure if the correct object does not lie in the retrieved top-k candidates.

\section{Dataset Details}
\label{sec:dataset}

\begin{figure}
\centering
    \includegraphics[width=0.90\linewidth]{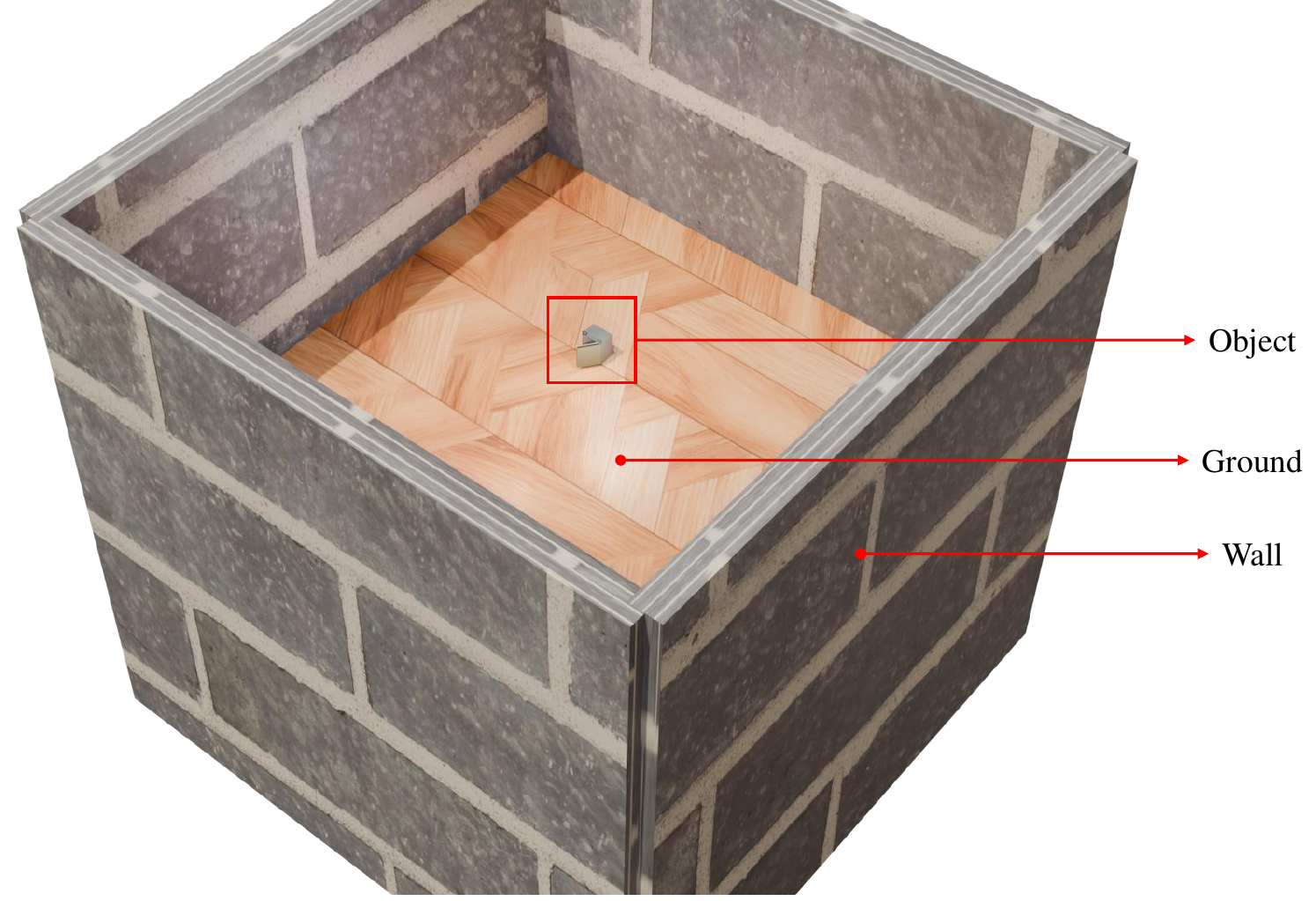}
    \caption{Illustration of the scene used to generate our synthetic dataset of articulating object videos.}
    \label{fig:scene_illustration}
\end{figure}

We build our dataset in the SAPIEN simulator~\cite{Xiang_2020_SAPIEN}. We design a small cuboid space with a simple texture as the environment. Then, we place the object in that space and render videos while manipulating the object using the built-in controller. \cref{fig:scene_illustration} illustrates the data collection environment. The statistics are listed in \cref{tab:data_config}.

\begin{table*}
    \centering
    \renewcommand{\arraystretch}{1.2}
    \begin{tabular}{lcccc}
        \toprule
        Category & \# Objects & Joint Type & \# Videos (revolute, prismatic) & \# Average Frames \\
        \midrule
        WashingMachine & 17 & revolute & 34 (34, 0) & 108.35 \\
        USB & 30 & revolute, prismatic & 60 (24, 36) & 149.5 \\
        StorageFurniture & 30 & revolute, prismatic & 120 (64, 56) & 122.8 \\
        Stapler & 8 & revolute & 20 (20, 0) & 118.9 \\
        Scissors & 30 & revolute & 60 (60, 0) & 91.56 \\
        Refrigerator & 25 & revolute & 76 (76, 0) & 97.18 \\
        Microwave & 16 & revolute & 32( 32, 0) & 134.31 \\
        Laptop & 30 & revolute & 60 (60, 0) & 126.03 \\
        Dishwasher & 29 & revolute & 58 (58, 0) & 136.93 \\
        Box & 19 & revolute & 50 (50, 0) & 129.08 \\
        Table & 50 & revolute, prismatic & 214 (32, 182) & 123.81 \\
        \midrule
        Total & 284 & revolute, prismatic & 784 (510, 274) & 121.68 \\
        \bottomrule
    \end{tabular}
    \caption{Dataset statistics. Our dataset contains a diverse set of object categories and is of significantly larger scale than prior work.}
    \label{tab:data_config}
\end{table*}

\section{Additional Evaluation and Discussion}
\label{sec:extra eval}

\subsection{Joint Estimation and Geometry Reconstruction Results on L-Dataset}
\label{subsec:eval_l}

\begin{table*}[t]
    \centering
    \renewcommand{\arraystretch}{1.2}
    \begin{tabular}{c l c c c c c}
        \toprule
        Joint & Methods
        & Axis (rad)$\downarrow$ & Position (m)$\downarrow$ & Type (\%)$\downarrow$ & State (rad or m)$\downarrow$ & Failure (\%)$\downarrow$ \\
        \midrule
        \multirow{5}{*}{Revolute}
          & \artany           & 0.73$\pm$0.77 & 0.73$\pm$0.40 & 37.12 & N.A. & 35.19 \\
          & \ours (Ours)                        & \textbf{0.29$\pm$0.51}  & \textbf{0.12$\pm$0.22} & \textbf{14.80} & \textbf{0.27$\pm$0.46} & 3.43 \\
          & \ours w/o Refine           & 1.09$\pm$0.62  & 0.24$\pm$0.33 & 73.60 & 0.59$\pm$0.51 & 2.57 \\
          & \ours w/o Coarse    & 1.30$\pm$0.25  & 0.34$\pm$0.23 & 76.82 & 0.68$\pm$0.44 & \textbf{1.07} \\
          & \ours w/o Segment    & 0.49$\pm$0.62  & 0.20$\pm$0.22 & 26.60 & 0.38$\pm$0.45 & 3.21 \\
        \cmidrule{1-7}
        \multirow{5}{*}{Prismatic}
          & \artany           & 0.83$\pm$0.78  & N.A. & 49.79 & N.A. & 44.89 \\
          & \ours                         & \textbf{0.27$\pm$0.40} & N.A. & \textbf{8.16} & \textbf{0.06$\pm$0.15} & 2.44 \\
          & \ours w/o Refine           & \textbf{0.27$\pm$0.44} & N.A. & 8.57  & 0.10$\pm$0.08 & \textbf{0.40} \\
          & \ours w/o Coarse    & 0.51$\pm$0.20  & N.A. & 14.69  & 0.12$\pm$0.15 & 1.63 \\
          & \ours w/o Segment    & 0.29$\pm$0.42  & N.A. & 9.79  & 0.08$\pm$0.15 & 2.44 \\
        \bottomrule
    \end{tabular}
    \caption{Evaluation of articulation parameter estimation on L-Dataset. We report the mean and standard deviation across all the test cases within the dataset. Our method achieves the lowest prediction error on most metrics.}
    \label{tab:L_results}
\end{table*}

\begin{table}[t]
    \centering
    \renewcommand{\arraystretch}{1.2}
    \resizebox{\columnwidth}{!}{
    \begin{tabular}{l c c c} 
        \toprule
        Methods & CD-w$\downarrow$ & CD-m$\downarrow$ & CD-s$\downarrow$ \\
        \midrule
         \artany & 0.08$\pm$0.21 & 0.42$\pm$0.64 & \textbf{0.04$\pm$0.15} \\
         \ours (Ours) & \textbf{0.01$\pm$0.03} & \textbf{0.10$\pm$0.21} & 0.06$\pm$0.19 \\
         \ours w/o Refine & \textbf{0.01$\pm$0.03} & 0.32$\pm$0.42 & 0.27$\pm$0.42 \\
         \ours w/o Coarse & \textbf{0.01$\pm$0.03} & 0.15$\pm$0.25 & 0.05$\pm$0.14 \\
         \ours w/o Segment & \textbf{0.01$\pm$0.03} & 0.35$\pm$0.43 & 0.30$\pm$0.44 \\
        \bottomrule
    \end{tabular}
    }
    \caption{Evaluation of geometry reconstruction on L-Dataset. We report the mean and standard deviation across all the test cases within the dataset. Our model is generally comparable to and superior to the baseline and ablated versions.}
    \label{tab:geometry_eval_l}
\end{table}

We conduct a more extensive evaluation of methods, excluding \rsrd, on the L-Dataset. Quantitative results on joint estimation and geometry reconstruction are shown in \cref{tab:L_results} and \cref{tab:geometry_eval_l}. Similar to the results in S-Dataset, our method outperforms \artany and other ablated versions consistently on L-Dataset.

\subsection{Variance of Joint Estimation Results}
\label{subsec:variance eval}

\begin{table*}[t]
    \centering
    \renewcommand{\arraystretch}{1.2}
    \begin{tabular}{c l c c c c c}
        \toprule
        Joint & Methods
        & Axis (rad)$\downarrow$ & Position (m)$\downarrow$ & Type (\%)$\downarrow$ & State (rad or m)$\downarrow$ & Failure (\%)$\downarrow$ \\
        \midrule
        \multirow{4}{*}{Revolute}
              & \ours (Ours)                         & \textbf{0.30$\pm$0.03} & \textbf{0.13$\pm$0.01} & \textbf{0.15$\pm$0.02} & \textbf{0.25$\pm$0.00} & 0.06$\pm$0.00 \\
          & \ours w/o Refine    & 1.15$\pm$0.08  & 0.26$\pm$0.01 & 0.75$\pm$0.06 & 0.63$\pm$0.04 & 0.04$\pm$0.00 \\
          & \ours w/o Coarse           & 1.26$\pm$0.01 & 0.34$\pm$0.00 & 0.79$\pm$0.03 & 0.69$\pm$0.01 & \textbf{0.02$\pm$0.00} \\
          & \ours w/o Segment    & 0.38$\pm$0.03  & 0.17$\pm$0.00 & 0.19$\pm$0.02 & 0.33$\pm$0.00 & 0.06$\pm$0.00 \\
        \cmidrule{1-7}
        \multirow{4}{*}{Prismatic}
              & \ours                         & \textbf{0.23$\pm$0.03} & N.A. & \textbf{0.06$\pm$0.03} & \textbf{0.05$\pm$0.01} & \textbf{0.00$\pm$0.00} \\
          & \ours w/o Refine           & 0.29$\pm$0.06  & N.A. & 0.11$\pm$0.04 & 0.10$\pm$0.02 & \textbf{0.00$\pm$0.00} \\
          & \ours w/o Coarse    & 0.48$\pm$0.01  & N.A. & 0.08$\pm$0.02 & 0.12$\pm$0.00 & 0.03$\pm$0.00 \\
          & \ours w/o Segment    & 0.25$\pm$0.03  & N.A. & 0.08$\pm$0.02 & 0.07$\pm$0.00 & \textbf{0.00$\pm$0.00} \\
        \bottomrule
    \end{tabular}
    \caption{Evaluating the variance of our method and ablated version. We run 10 experiments with different seeds on S-Dataset. We compute the mean value of each run and report the mean and standard deviation of these mean values. Our method consistently outperforms the ablated version of our method.}
    \label{tab:variance_results}
\end{table*}

To further evaluate the stability of our method, we conduct experiments on the S-Dataset with 10 different seeds. The mean and standard deviation of the final results are shown in \cref{tab:variance_results}. We can find that our method consistently outperforms the ablated version of our method. Besides, by comparing our results and ours w/o Refinement results, we can find that the refinement module can reduce the variance of the prediction.

\subsection{Moving Map Segmentation Results}
\label{subsec:moving map eval}

\begin{table}
    \centering
    \renewcommand{\arraystretch}{1.2}
    \begin{tabular}{lc} 
        \toprule
        Methods & mIOU$\uparrow$ \\
        \midrule
        \ours (Ours) & 0.59653 \\
        \ours w/o Refine (MonST3R) & \textbf{0.61632} \\
        \ours w/o Coarse & 0.42545 \\
        \ours w/o Segment & 0.40310 \\
        \bottomrule
    \end{tabular}
    \caption{Evaluation on moving map. Our method is slightly worse than MonST3R. Our method will filter out newly observed parts, which could be moving parts in the video. Thus, the moving map estimation is slightly worse.}
    \label{tab:segment_eval}
\end{table}

Moving map segmentation is an important intermediate results of our method. We evaluate moving map segmentation using mIOU metric. From the quantitative results in \cref{tab:segment_eval}, we find our method achieves comparable results to MonST3R~\cite{zhang2024monst3r}, though our pipeline ignores the inner parts of the object, which could be moving part in the video. Without coarse prediction as initialization, our method is hard to optimize for very accurate moving map segmentation. Similarly, without automatic mask generation, our pipeline cannot optimize moving map of the video efficiently, resulting in worse moving map segmentation of the video.

\subsection{Camera Pose Estimation Results}
\label{subsec:cam pose eval}

\begin{table}
    \centering
    \renewcommand{\arraystretch}{1.2}
    \begin{tabular}{lcc} 
        \toprule
        Methods & Rot. Error(rad)$\downarrow$ & Pos. Error(m)$\downarrow$ \\
        \midrule
        MonST3R & 0.14482 & \textbf{0.08729} \\
        CUT3R & 0.14663 & 0.10536 \\
        \ours (Ours) & \textbf{0.08066} & 0.09246 \\
        \ours w/o Refine & 0.11814 & 0.10918 \\
        \ours w/o Coarse & 0.11073 & 0.16544 \\
        \ours w/o Segment & 0.11547 & 0.16892 \\
        \bottomrule
    \end{tabular}
    \caption{Evaluation on camera pose estimation. Our method produces accurate camera pose estimation, while other methods have larger error on either rotation or position estimation.}
    \label{tab:camera_eval}
\end{table}

\begin{table}
    \centering
    \renewcommand{\arraystretch}{1.2}
    \begin{tabular}{lcc} 
        \toprule
        Dataset & Rot. Error(rad)$\downarrow$ & Pos. Error(m)$\downarrow$ \\
        \midrule
        S-Dataset & 0.008 & 0.020 \\
        L-Dataset & 0.011 & 0.029 \\
        \bottomrule
    \end{tabular}
    \caption{Evaluation on aligning the camera pose of the first video frame to the static scan coordinate system. We test our simple strategy on both the S-Dataset and L-Dataset. The results show that this simple aligning strategy used in our real-world experiment is reliable enough.}
    \label{tab:align_eval}
\end{table}

There are already several dynamic scene reconstruction works that can infer camera poses of a video sequence. This comparison aims to demonstrate why not directly use camera estimation results from existing dynamic scene reconstruction models. We compare our camera pose prediction results with two very recent works MonST3R~\cite{zhang2024monst3r} and CUT3R~\cite{cut3r}. We transform all the camera poses to the camera coordinate of the first video frame. We compare both camera rotation error and translation error on our dataset. From the results shown in \cref{tab:camera_eval}, we can find that our method can produce more accurate camera pose estimation, particularly camera orientation, against recent works on dynamic scene understanding. We further evaluate how camera pose estimation will affect the final joint parameter estimation results on S-Dataset. In \cref{tab:monst3rcam_results}, we find that using camera poses estimated by MonST3R will lead to a large degradation in the performance.

We also evaluate the accuracy of the strategy of aligning the first video frame to the static scan coordinate system in \cref{tab:align_eval}. On both datasets, the camera pose estimation error is small, showing the effectiveness of our strategy of aligning the first video frame to the scan coordinate system.

\begin{table*}[t]
    \centering
    \renewcommand{\arraystretch}{1.2}
    \begin{tabular}{c l c c c c c}
        \toprule
        Joint & Methods
        & Axis(rad)$\downarrow$ & Position(m)$\downarrow$ & Type(\%)$\downarrow$ & State(rad or m)$\downarrow$ & Failure(\%)$\downarrow$ \\
        \midrule
        \multirow{2}{*}{Revolute}
              & \ours (Ours)                         & \textbf{0.32$\pm$0.56} & \textbf{0.13$\pm$0.25} & \textbf{15.90} & \textbf{0.25$\pm$0.46} & 6.81 \\
          & \ours w/ MonST3R camera    & 0.74$\pm$0.58  & 0.27$\pm$0.25 & 51.11 & 0.39$\pm$0.43 & \textbf{6.67} \\
        \cmidrule{1-7}
        \multirow{2}{*}{Prismatic}
              & \ours                         & \textbf{0.24$\pm$0.33} & N.A. & 10.34 & \textbf{0.08$\pm$0.22} & 0 \\
          & \ours w/ MonST3R camera          & 0.68$\pm$0.31  & N.A. & \textbf{3.33} & 0.15$\pm$0.10 & 0 \\
        \bottomrule
    \end{tabular}
    \caption{We test how camera poses will affect the joint parameter estimation on S-Dataset. We can find that using camera poses estimated by MonST3R will lead to a large degradation in the final performance.}
    \label{tab:monst3rcam_results}
\end{table*}

\subsection{Moving Map Prediction Choice}
\label{subsec:moving map choice}

\begin{figure*}
\centering
    \includegraphics[width=0.95\textwidth]{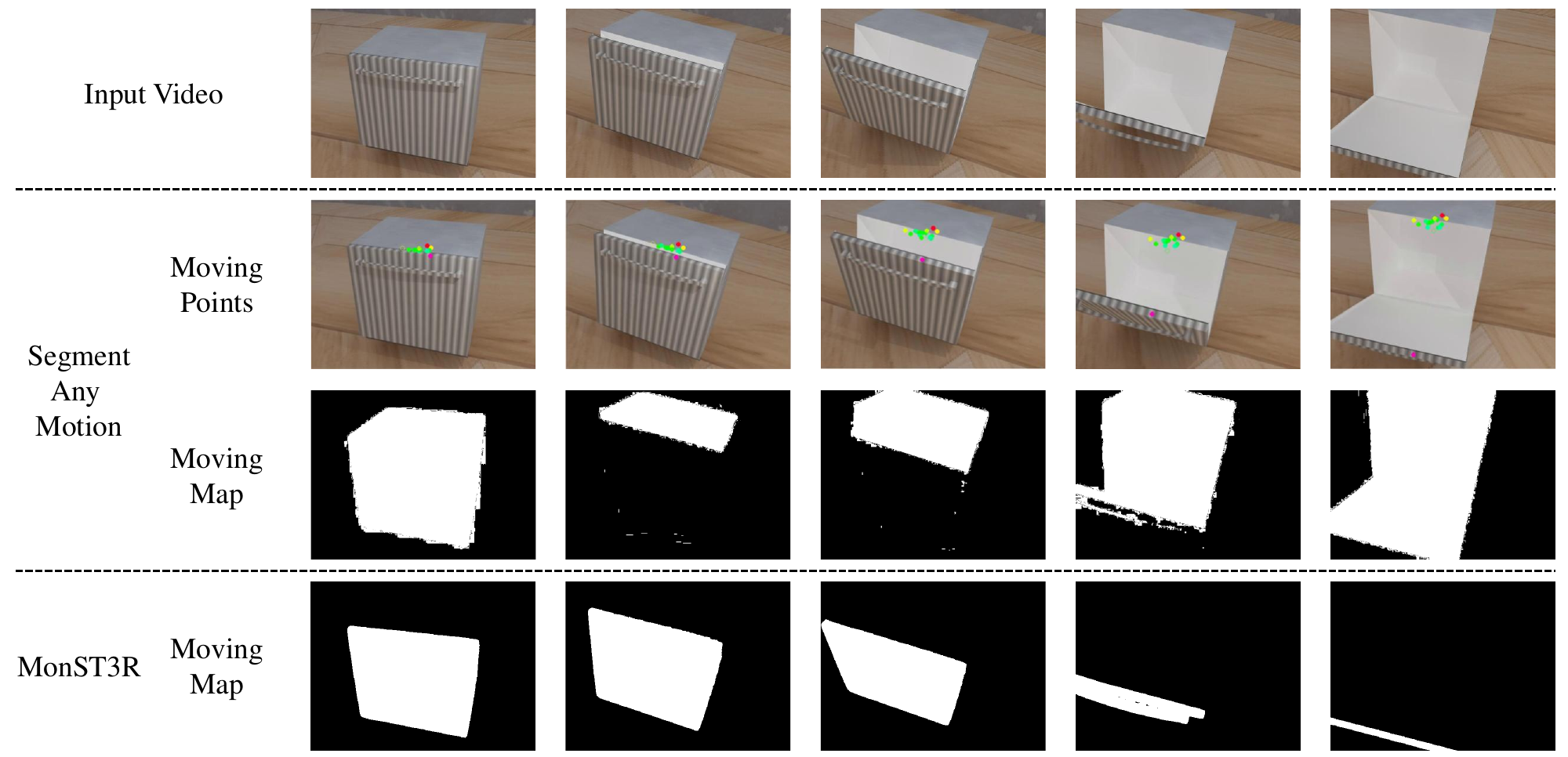}
    \caption{Comparison of moving map prediction between MonST3R and Segment Any Motion. From this example, we can find that the point-track-based method is very sensitive to the location of the points. Moving points predicted by Segment Any Motion lie on the edge of the door of this dishwasher. Using these points as a prompt for Segment Anything 2~\cite{ravi2024sam2} will likely segment the wrong parts and objects.}
    \label{fig:moving map}
\end{figure*}

In our coarse prediction module, we use MonST3R~\cite{zhang2024monst3r} to generate a moving map of the input video. A natural question is, why not use an off-the-shelf motion segmentation method? Here we compare the output of moving map prediction results from MonST3R and Segment Any Motion~\cite{huang2025segmentmotionvideos}, which is a very recent work on motion segmentation on videos. In \cref{fig:moving map} we can find that Segment Any Motion produces much larger error prediction results compared to MonST3R. Segment Any Motion first computes point tracks of the video, and selects moving points based on point tracks. Then, it uses these points to prompt Segment Anything 2~\cite{ravi2024sam2} to generate the moving map of the video~\cite{huang2025segmentmotionvideos}. But in this case, the moving points lie around the edge of the door of this dishwasher. Thus, some points prompt selecting the dishwasher's top as the moving part instead of the front door. This small position error will lead to a large error in moving map segmentation. In contrast, MonST3R computes the moving map based on optical flow and estimated camera poses. It subtracts the optical flow caused by camera motion from the final optical flow to determine which part of the video is moving. This strategy turns out to be much robust to point-track based motion segmentation methods.

\subsection{Discussion on Automatic Video Segmentation Strategy}

\begin{figure*}
\centering
    \includegraphics[width=0.95\textwidth]{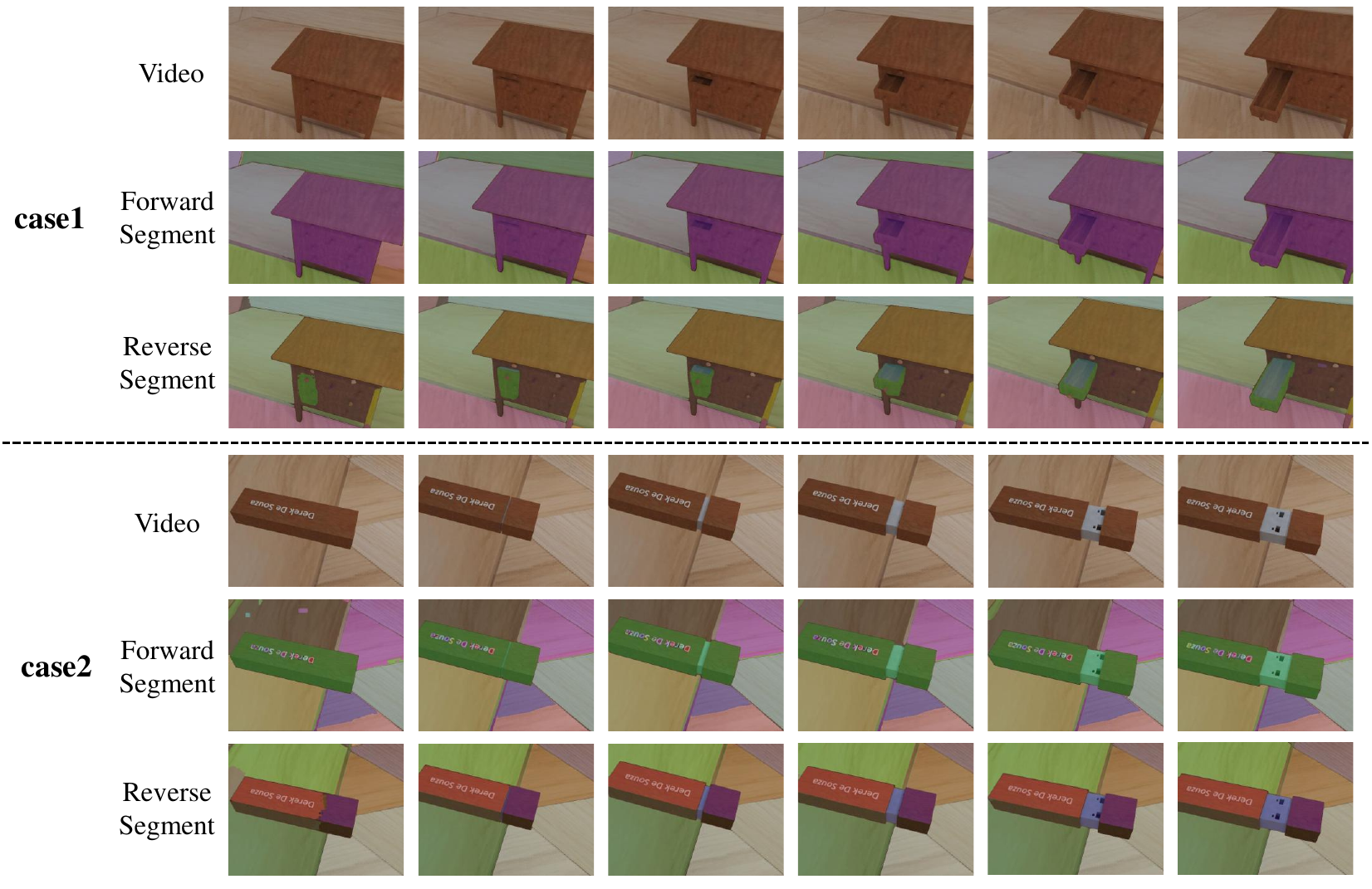}
    \caption{Examples of automatic part segmentation in forward and reverse temporal directions. From these two cases, we see that segmenting in the reverse direction makes it easier to identify different parts of the object, particularly when the surface texture is similar.}
    \label{fig:autoseg-strategy}
\end{figure*}

Automatic part segmentation on the video plays a vital role in our pipeline. We find that automatic part segmentation performs better on the reversed video. Most videos in our dataset are about opening the door or drawer. In this case, some movable parts are difficult to segment out due to the very high similarity of texture of every part of the object. But when the door is opened or the drawer is pulled out, it's much easier to distinguish the movable part from the other parts of the object. \cref{fig:autoseg-strategy} illustrates the segmentation results on a normal and reversed video.

Besides, we test our method using ground truth mesh-level segmentation on the input video on S-Dataset. An image is rendered from an object comprising several meshes. The mesh-level segmentation identifies different regions that are rendered from different meshes. The results in \cref{tab:meshseg_results} show that at least using automatic part segmentation is comparable to using ground truth mesh-level segmentation.

\begin{table*}[t]
    \centering
    \renewcommand{\arraystretch}{1.2}
    \begin{tabular}{c l c c c c c}
        \toprule
        Joint & Methods
        & Axis(rad)$\downarrow$ & Position(m)$\downarrow$ & Type(\%)$\downarrow$ & State(rad or m)$\downarrow$ & Failure(\%)$\downarrow$ \\
        \midrule
        \multirow{2}{*}{Revolute}
              & \ours (Ours)                         & \textbf{0.32$\pm$0.56} & \textbf{0.13$\pm$0.25} & \textbf{15.90} & 0.25$\pm$0.46 & 6.81 \\
          & \ours w/ mesh-level seg    & 0.34$\pm$0.58  & 0.15$\pm$0.27 & 17.77 & \textbf{0.20$\pm$0.40} & \textbf{6.67} \\
        \cmidrule{1-7}
        \multirow{2}{*}{Prismatic}
              & \ours                         & 0.24$\pm$0.33 & N.A. & 10.34 & 0.08$\pm$0.22 & 0 \\
          & \ours w/ mesh-level seg          & \textbf{0.21$\pm$0.36}  & N.A. & \textbf{10.00} & \textbf{0.03$\pm$0.13} & 0 \\
        \bottomrule
    \end{tabular}
    \caption{We compare joint parameter estimation results using different automatic video segmentation strategies on S-Dataset. Our automatic part segmentation is comparable to using ground truth mesh-level segmentation.}
    \label{tab:meshseg_results}
\end{table*}

\begin{figure*}
\centering
    \includegraphics[width=\textwidth]{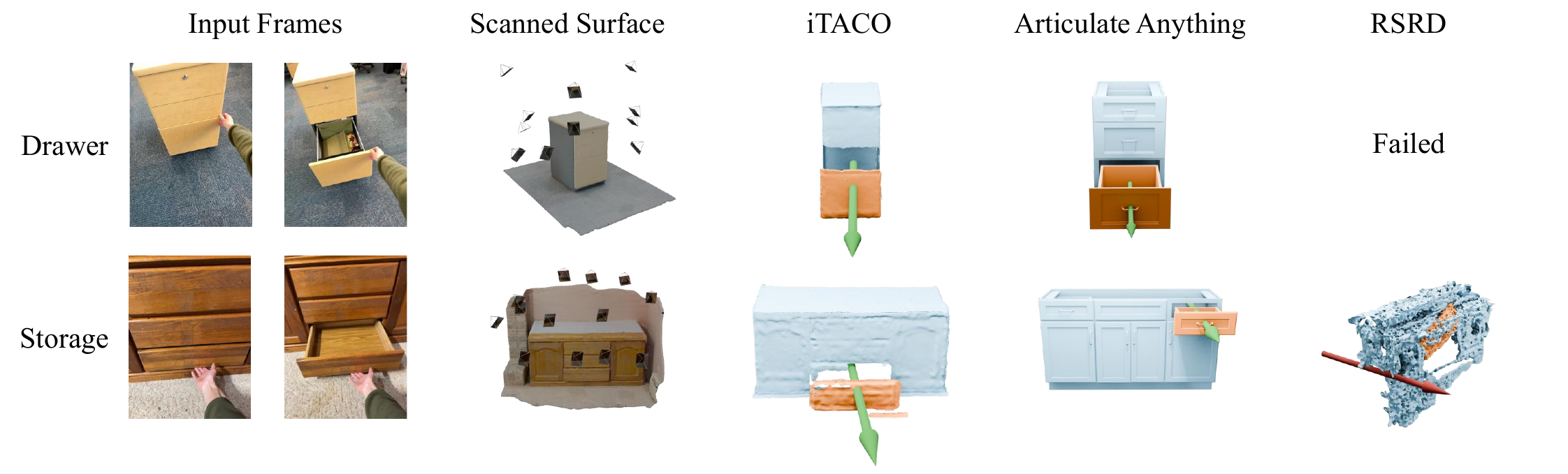}
    \caption{Qualitative results on real data.}
    \label{fig:real_qualitative_result2}
\end{figure*}

\subsection{Additional Real-World Results}
\label{subsec:real obj eval}

In the main paper, we demonstrate the results of building interactable digital twins of real articulated objects with revolute joints. Here we demonstrate reconstructing articulated objects with prismatic joints in \cref{fig:real_qualitative_result2}. \artany still struggles with retrieving the correct object in the Storage case, while \rsrd fails to crop and cluster the object from the environment for the Drawer case.

\begin{figure*}
\centering
    \includegraphics[width=\textwidth]{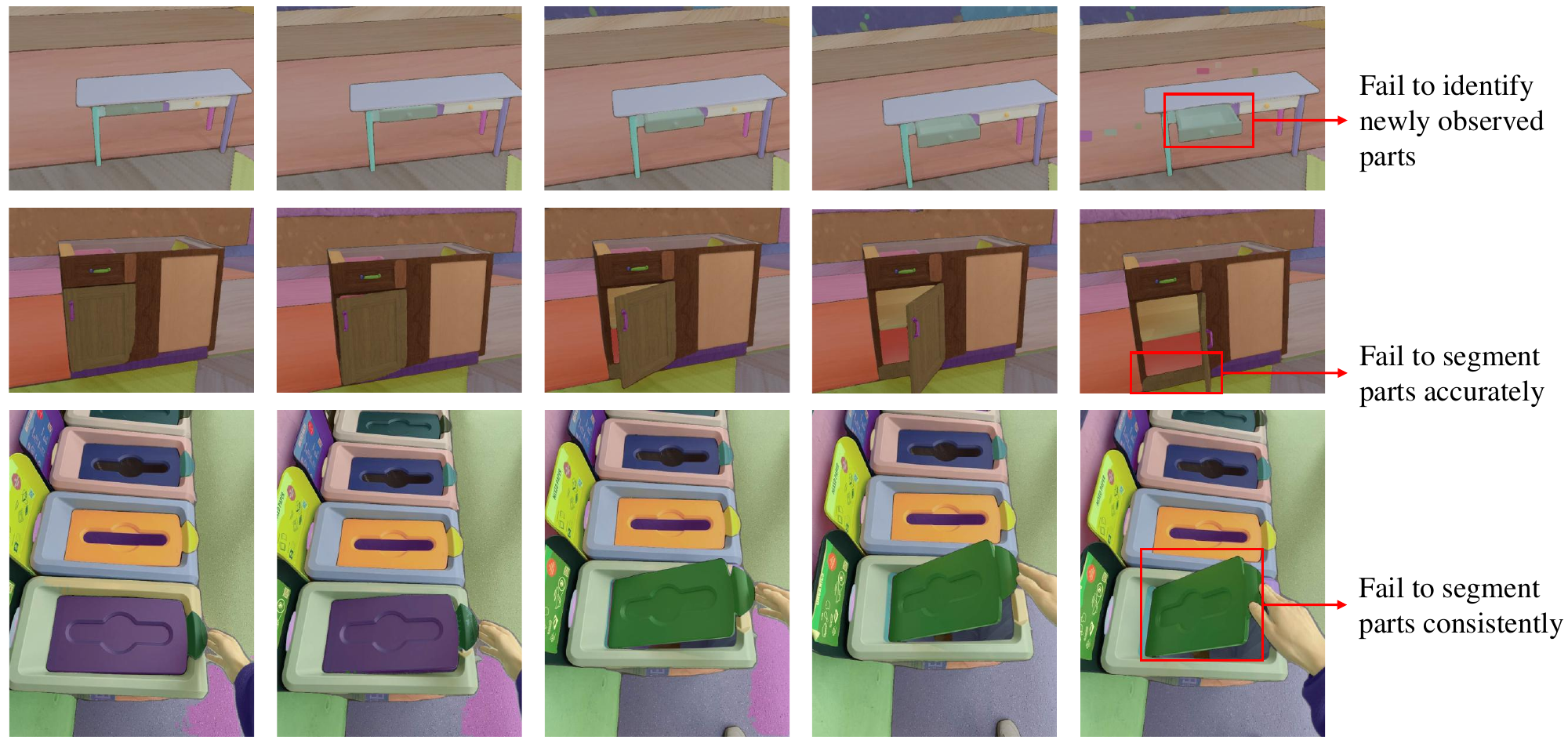}
    \caption{Three common failure modes observed during automatic part segmentation.}
    \label{fig:failure mode}
\end{figure*}

\subsection{Failure Case Analysis and Limitation}
\label{subsec:failure case}

Our pipeline relies heavily on automatic part segmentation, since the refinement module is built upon part representation. We summarize three common failure modes in \cref{fig:failure mode}. The first row shows that auto part segmentation may not identify newly observed parts in the video. It combines the interior of the drawer with its front face. When computing the Chamfer distance for optimization, the interior parts cannot find correspondence on the object surface points cloud, resulting in inaccurate estimation. The second row shows that auto part segmentation may not segment parts accurately. In this case, a part of the storage body is segmented out with the door. This part is static while the door is moving. This will also introduce noise to the Chamfer distance. The third row illustrates that auto part segmentation may not segment parts consistently. The lid of the trash can belongs to different segments in the video, shown in two different colors. Our pipeline will filter out parts that do not appear in the first frame. In this case, the trash can lid will be considered as newly observed parts and filtered out in the later sequence of the video, losing important information for joint estimation. 

From the failure modes discussed above, we believe geometry information will help to refine the segmentation results. For example, we can use recent 3D segmentation approaches like SAMPart3D~\cite{yang2024sampart3d} and PartField~\cite{partfield2025} to produce segmentation on the observed point cloud. This geometry-based segmentation can correct the errors in automatic part segmentation, which is purely image-based.

\end{document}